\documentclass[a4paper,11pt]{article}

\usepackage{longtable}
\usepackage[fleqn]{amsmath}
\usepackage{amssymb,amsthm}
\usepackage{graphicx}
\usepackage{bm}
\usepackage{footmisc}
\usepackage{afterpage}
\usepackage{float}
\usepackage{lscape}
\usepackage{longtable}

\usepackage{braket}
\usepackage{comment}

\usepackage[square,comma,sort&compress,numbers]{natbib}

\usepackage[dvipsnames]{xcolor}
\definecolor{gray75}{gray}{0.4}

\usepackage{fullpage}
\usepackage{etex}
\usepackage[utf8]{inputenc}
\usepackage[T1]{fontenc}
\usepackage{lmodern}
\usepackage[nolist,nohyperlinks]{acronym}
\usepackage{booktabs}
\usepackage{hepunits}
\usepackage{units}

\usepackage{multirow}

\usepackage{epigraph}
\epigraphsize{\small\itshape}
\usepackage{csquotes}

\usepackage{slashed}
\usepackage[section]{easy-todo}
\usepackage{caption}
\usepackage{subcaption}
\usepackage{setspace}
\usepackage{notoccite}
\usepackage{listings}
\usepackage{enumerate}

\usepackage[hyperfootnotes=false,pdfpagelabels]{hyperref}  

\hypersetup{%
    colorlinks=true, linktocpage=true, pdfstartpage=3, pdfstartview=FitV,%
    breaklinks=true, pdfpagemode=UseNone, pageanchor=true, pdfpagemode=UseOutlines,%
    plainpages=false, bookmarksnumbered, bookmarksopen=true, bookmarksopenlevel=1,%
    hypertexnames=true, pdfhighlight=/O,
   urlcolor=gray75, linkcolor=gray75, citecolor=gray75,
     pdftitle={Quantization and time evolution},%
     pdfauthor={Leonardo Antunes Pedro},%
}

\usepackage{mciteplus}

\newcommand*{\bea}{\begin{eqnarray}}
\newcommand*{\eea}{\end{eqnarray}}
\newcommand*{\be}{\begin{equation}}
\newcommand*{\ee}{\end{equation}}

\newcommand*{\tr}{\mathrm{tr}}

\newcommand{\bma}{\begin{pmatrix}}
\newcommand{\ema}{\end{pmatrix}}

\graphicspath{{./Images/}}

\renewcommand{\textflush}{flushepinormal}

\theoremstyle{plain}

\newtheorem*{rmk*}{Note}

\theoremstyle{definition}

\newtheorem*{defn*}{Definition}

\theoremstyle{remark}

\makeatletter
\renewcommand{\@epitext}[1]{
\itshape \begin{minipage}{\epigraphwidth}\begin{\textflush} #1
\end{\textflush}\end{minipage}\vspace{1ex}}

\makeatother
\title{Quantization due to the time evolution, with applications to Quantum Yang-Mills theory, Quantum Gravity and Classical Statistical Field Theory}
\author{Leonardo Pedro\\
  leonardo@cftp.tecnico.ulisboa.pt}
\date{\today}
\begin{document}
\maketitle

\begin{abstract}
Quantum Yang-Mills theory, Classical Statistical Field Theory (for Hamiltonians which are non-polynomial in the fields, e.g. General relativistic statistical mechanics) and Quantum Gravity all suffer from severe mathematical inconsistencies and produce unreliable predictions at best.

We define with mathematical rigor, a class of statistical field
theories in Minkowski space-time where the (classical)
canonical coordinates when modified by a non-deterministic time evolution,
verify the canonical commutation relations. We then extend these statistical
field theories to include non-trivial gauge symmetries and show that these theories
have all the features of a Quantum Yang-Mills theory in four-dimensional space-time. We generalize the Gaussian measure to allow for the definition of Hamiltonians which are non-polynomial in the fields, such as in Classical Statistical Field Theory and Quantum Gravity.
Finally, we test the consistency of our formalism with the quantization of the free Electromagnetic field.
\end{abstract}

\section{Introduction}


Schr\"odinger described quantization as the consequence of solving an eigenvalue problem for the Hamiltonian~\cite{qmeigenvalue}: in an infinite-dimensional linear space of functions, continuous and
discrete (i.e. quantized) energy spectra may coexist.
Thus, from the very beginning there was a relation between the time evolution (defined by the Hamiltonian) and the notion of quantization.

There is no doubt that the best known description of the experimental
data collected so far is based on a quantum
theory~\cite{pdg}. However, the notion of quantization is not much
clearer than it was in 1926~\cite{mistery}. Up to now there is no definition or an example of a Quantum Yang-Mills theory in four-dimensional space-time, with mathematical rigor and in the absence of approximations~\cite{prize}. The status of Quantum Gravity is even worse, since no reliable way to make predictions is known~\cite{pullin2020loop}. Classical Statistical Field Theory currently depends on the same renormalization formalism of Quantum Yang-Mills theory~\cite{mussardosft}, without mathematical rigor for Hamiltonians which are non-polynomial in the fields (e.g. General relativistic statistical mechanics~\cite{Rovelli:2012nv}) and a peculiar notion of ``continuum'' (see Section~\ref{sec:measure}). In this paper we will
change this status, proposing a simple and mathematically meaningful
definition of quantization with applications to Quantum Yang-Mills theory, Quantum Gravity and Classical Statistical Field Theory.

For comparison purposes, we now address what quantization is not.
Quantization is not replacing Poisson bracket's by  canonical
commutation relations. The method of replacing Poisson bracket's by
canonical commutation relations can always be applied
(for analytical functions), it is called
prequantization~\cite{hall2013quantum,functor}.
However it doesn't lead by itself to useful results (hence the name
prequantization). Of course, we can try to improve the method  so that it leads to useful results (this is the geometrical quantization program~\cite{hall2013quantum}),
however we end up with a definition of quantization which is so complex and arbitrary that it is not useful in practice, in particular in the presence of gauge symmetries. 

Quantization is also not second quantization (based on a Fock-space), which relates a quantum description of a single-particle system to a quantum description of a many-particle system~\cite{functor}. We can only apply second quantization to a quantum theory, hence the name ``second''.

Quantization is also not computing the Feynman's path integral, since we know that the Feynman's path integral does not have the property (sigma-additivity) which allows
computation of the integral by approximating the
integrand~\cite{nonperturbativefoundations}, and thus it is not an
integral. Of course as in prequantization, we can try to improve the path integral~\cite{mathematicalpath}, however we are very far from a consistent definition of path integral which is useful in practice.

Quantization is also not a perturbative expansion or a lattice regularization. These two different approximations are useful and have a clear definition, but since we know that they are complementary~\cite{nonperturbativefoundations} then neither of them can be used to define quantization.

Note that there is enough experimental evidence to conclude that all the methods above mentioned---namely prequantization, second quantization, Feynman's path integral, perturbative expansion, lattice reguralization---are related to the quantum phenomena and thus they are necessarily related with the definition of quantization. But we insist that it is also clear that none of them by itself can be used to define quantization.

There is a big conceptual problem with the notion of quantization: we are trying to relate a deterministic theory (classical mechanics) with a non-deterministic theory (quantum mechanics). From the point of view of (classical) information theory~\cite{info}, the root of probabilities (i.e. non-determinism) is the absence of information. Statistical methods are required whenever we lack complete information about a system, as so often occurs when the system is complex~\cite{bertinstatistical}. Thus we can convert a deterministic theory to a statistical theory unambiguously (using trivial probability distributions); but we cannot convert a statistical theory into a deterministic theory unambiguously since we need new information~\footnote{E.g. the assumptions required by the deterministic models in reference~\cite{automaton} are new information.}.

On the other hand, the relation between quantum mechanics and a
statistical theory (both are non-deterministic) is clear: the wave-function is nothing else than one possible parametrization of any probability distribution~\cite{parametrization}; the parametrization is a surjective map from an hypersphere to the set
of all possible probability distributions. Two wave-functions are always related by a rotation of the hypersphere, which is a linear transformation and it preserves the hypersphere. It is thus a good parametrization which allows us to represent a group of symmetry transformations (such as translations in space and time
or rotations in space) using linear transformations of the hypersphere. These symmetry transformations may be non-deterministic, thus quantum mechanics is a generalization of classical statistical mechanics (but not of probability theory, thanks to the wave-function's collapse~\cite{parametrization}).

The non-commutativity of operators is thus intrinsic to any statistical theory. This saves us from the need to ``deform'' commutative algebras into non-commutative ones upon quantization~\cite{Berra-Montiel:2020egf}.
In our opinion, either the quantization of a classical theory or the classical limit of a quantum theory cannot go much beyond Koopman-von Neumann version of classical mechanics~\cite{Sudarshan1976}, i.e. a description of classical mechanics as a statistical theory (which is always possible, since a deterministic theory is a particular case of a statistical theory).


In Section~\ref{sec:gauge} we will define gauge symmetry exclusively within the Hamiltonian formalism (that is, a definition independent from the Lagrangian formalism);  in Section \ref{sec:constraints} we will relate constraints with gauge symmetries; in Section~\ref{sec:lorentz} we will show how to deal with Lorentz covariance within the Hamiltonian formalism;
in Section~\ref{sec:source} we will define a Statistical Source Field Theory where the fields in phase-space also depend on a time coordinate; in Section~\ref{sec:time} we show that the canonical commutation relations between momentum and position appear due to a non-deterministic time-evolution; in Section~\ref{sec:gaussian} we study the implications of using a gauge-variant gaussian measure for the phase diagram; 
in Section~\ref{sec:local} we discuss how to define operators which are effectively local, despite respecting the momentum constraint; in Section~\ref{sec:brst} we discuss the relation of our definition of Quantum Gauge Field Theory with the BRST formalism; in Section~\ref{sec:rg} we discuss Renormalization, the mass gap and the status of the problem from the Millennium prize; in Sections~\ref{sec:ym}~and~\ref{sec:em} we apply our formalism to the Yang-Mills and free electromagnetic field and the results are compatible with the existing literature; up to Section~\ref{sec:measure} we assume that there is no spontaneous symmetry breaking, where we generalize the Gaussian measure to allow for the definition of Hamiltonians which are non-polynomial in the fields, such as in Classical Statistical Field Theory and Quantum Gravity; in Section~\ref{sec:spontaneous} we finally discuss spontaneous symmetry breaking, but already within the framework of the measure defined in the previous section which allows for a more clear discussion than with the Gaussian measure; in Section~\ref{sec:gravity} we show how diffeomorphisms can be incorporated within our formalism and we propose an Hamiltonian for quantum gravity; finally in Section~\ref{sec:conclusion} we end with the Conclusions.

\section{Gauge symmetry: definition within Quantum Field Theory}
\label{sec:gauge}

The concept of gauge symmetry is clear in classical field theory: it is a particular symmetry of a singular Lagrangian. A singular Lagrangian is a Lagrangian which does not determine the time-evolution of all fields. Consequently, there are fields whose time-evolution is arbitrary and so these fields cannot be physical fields. The gauge symmetries are the symmetries of the Lagrangian which transform the fields with one time-evolution to fields with another time-evolution. The singular Lagrangian only determines the time-evolution of the gauge-invariant algebraic combinations of fields, thus only these  algebraic combinations are observable.

However, the  gauge symmetry clashes at a fundamental level with the canonical commutation relations of Quantum Field Theory.

If the Lagrangian is singular, then in the corresponding classical Hamiltonian formalism there are fields whose conjugate momenta is constrained to be null. But this is incompatible with the canonical commutation relations of the corresponding quantum fields: the commutator of a field with its conjugate momentum cannot be null.


The way out is to reformulate the Hamiltonian formalism of a classical gauge theory. The classical Dirac brackets and the associated Dirac's constrained Hamiltonian formalism are a previous attempt, where only some variables are ``physical'' and thus are part of the corresponding Quantum theory. This would imply that a gauge theory can only be defined within the classical Lagrangian formalism, and then the canonical commutation relations can only be defined after the elimination of the ``non-physical'' variables which necessarily implies the elimination of the gauge symmetry (i.e. gauge-fixing, at least for the part of the gauge symmetry which is the source of the ``non-physical'' variables); it would imply that there is not really a fully quantum gauge theory, only a quantum theory without the full gauge symmetry. Moreover, the elimination of the ``non-physical'' variables is also fraught with severe technical problems such as the Gribov
ambiguity~\cite{henneaux1992quantization}. 


The use of classical Dirac brackets as a step towards a quantum gauge theory was indeed an unfortunate attempt. Since the wave-function is a parametrization of any probability distribution, in fact the canonical commutation relations of Quantum Field Theory should be used as a step towards an Hamiltonian formulation of classical gauge theory (and also of quantum gauge theory, of course) and not the other way around. At the cost of using non-commutative operators (and thus of using a non-deterministic formulation), we get the crucial advantage that the phase-space \emph{coincides} with the coordinate  space, i.e. every coordinate comes equipped with a conjugate momentum operator \emph{without} enlarging the phase-space. Note that the conjugate momentum operator is not part of the phase-space in this case, since it does not commute with the corresponding coordinate.

Once we choose to use the canonical commutation relations, the remaining question is merely what is the correct phase-space for our problem. In Source Field Theory (defined in Section~\ref{sec:source}), the source fields do have a time coordinate and are also part of the phase-space, so that the Lagrangian becomes dispensable. We will continue discussing the definition of gauge symmetry within Quantum Field Theory in sections~\ref{sec:constraints} and~\ref{sec:brst}.

Note that a formulation of classical gauge theories within a Quantum formalism
(besides classical statistical and quantum gauge theories), may also be advantageous.
Theories such as classical electrodynamics or more generally classical non-abelian gauge theories~\cite{classicalsolutions} involve a system of non-linear partial differential equations. It is a very hard problem to study in general the space of classical solutions of such systems\footnote{A well known example is the Navier-Stokes equation~\cite{navierstokes,*navierstokes2}.}. Even when a few solutions can be found, they may not be the ones that describe the physical system correctly. A consistent theory  covering many cases only exists (at the moment) for systems of linear partial
differential equations~\cite{ciarlet2013linear}. Thus to solve many
non-linear deterministic theories we may not have better alternative
(at the moment) than to consider them as a particular case of a
statistical theory and apply linear quantum methods on its wave-function parametrization~\cite{Sudarshan1976,rkhs}---then the building blocks of the overall deterministic theory are non-deterministic.

\section{Constraints and gauge symmetry}
\label{sec:constraints}

In the Hamiltonian formalism, the constraints are from a technical point of view, a representation of an ideal by the zero number. By an ideal we mean an ideal in the algebraic sense. We assume here the algebra of operators to be non-commutative, so that we can cover Quantum theory.

The correspondence between geometric spaces and commutative algebras is important in algebraic geometry\footnote{The correspondence between geometric spaces and commutative algebras is consequence of the Gelfand representation: there is an isomorphism between a commutative C*-algebra $A$ and the algebra of continuous functions of the spectrum of $A$.}. It is usually argued that the phase space in quantum mechanics corresponds to a non-commutative algebra and thus it is a non-commutative geometric space in some sense~\cite{connesnoncommutative}.
However, after the wave-function collapse, only a commutative algebra of operators remains~\cite{parametrization}. Thus, the phase space in quantum mechanics is a standard geometric space and the standard spectral theory (where the correspondence between geometric spaces and commutative algebras plays a main role~\cite{spectralhistory}) suffices.


It suffices to constrain to zero the Casimir operators of the algebra of constraints, these are included in the commutative algebra (for abelian and also for non-abelian symmetries). This imposes the constraints without the need for the constraints to be part of the commutative algebra.

Once non-determinism is taken into account, then non-commutative operators can be taken into account and the constraints are the generators of a gauge symmetry group. In case the Lie group is infinite-dimensional, there is some ambiguity in its definition~\cite{infinitelie,infinitelie}. Since the Hilbert space is a representation of the gauge symmetry, we consider instead the $C^*$-algebra~\cite{realoperatoralgebras} generated by the unitary operators on an Hilbert space of the form $e^{i\int d^4 x \theta(x) G(x)}$ where $G(x)$ is a constraint and $\theta(x)$ is a square integrable function of space-time $x$.

Note that the algebra of observable operators already conserves the constraints (i.e. it is a trivial representation of the gauge symmetry), so the Hilbert space does not need to verify the constraints (i.e. it may be a non-trivial representation of the gauge symmetry). In fact, in many cases it would be impossible for the cyclic state of the Hilbert space to verify the constraints, as it was noted long ago:

\epigraph{``So we have the situation that we cannot define accurately the vacuum state. We therefore have to work with a standard ket $|S>$ which is ill-defined. One can, however, do many calculations without using the accurate conditions \emph{[vacuum verifies constraints]} and the successes of quantum electrodynamics are obtained in this way.''}{Paul Dirac (1955)~\cite{Dirac:1955uv}}

Indeed, there are some symmetries of the algebra of operators which necessarily the expectation functional  cannot have, since the expectation functional is a trace-class operator (the expectation of the operator 1 is 1) and its dual-space is bigger (the space of bounded operators). 

For instance, consider an infinite-dimensional discrete basis $\{e_k\}$ of an Hilbert space (indexed by the integer numbers $k$) and the symmetry group generated by the transformation $e_k\to~e_{k+1}$ (translation). There is no normalized wave-function (and thus no expectation functional) which is translation-invariant, while there is a translation-invariant algebra of bounded operators (starting with the identity operator). 


We define gauge-fixing as comprehensive whenever it crosses all possible gauge-orbits at least once. On the other hand, we define gauge-fixing as complete whenever it crosses all possible gauge-orbits at most once, i.e. when there is no remnant gauge symmetry. The Dirac brackets require the gauge-fixing to be both comprehensive and complete, which is not possible in general due to the Gribov ambiguity~\cite{henneaux1992quantization}. 
In a non-abelian gauge-theory, the Gribov ambiguity forces us to consider a phase-space formed by fields defined on not only space but also time. This is related to the fact that in a fibre bundle (the mathematical formulation of a classical gauge theory) the time cannot be factored out from the total space because the topology of the total space is not a product of the base-space (time) and the fibre-space, despite that the total space is \emph{locally} a product space.
Thus, the Hamiltonian constraints cannot be interpreted literally, that is, as mere constraints in a too large phase-space whose ``non-physical'' degrees of freedom need to be eliminated. Moreover, this picture makes little sense in infinite-dimensions: the gauge potentials can be fully reconstructed from the algebra of gauge-invariant functions, apart from the gauge potential at a neighborhood of one specific arbitrary point in space-time~\cite{wilsonloops}.

If we consider instead a commutative algebra with a spectrum, such that any non-trivial gauge transformation necessarily modifies the spectrum (e.g. the gauge field $A_\mu$ which is a function of space-time), then one point in the spectrum is one example of a complete non-compreehensive gauge-fixing. The gauge-fixing is non-comprehensive because the action of the gauge group on the spectrum is not transitive. Such commutative algebra has the crucial advantage that the constraints are necessarily excluded from the algebra, so that it can be used to construct a standard Hilbert space which is compatible with the constraints because the relevant operators of the commutative algebra are the ones commuting with the constraints, saving us the need to eliminate the ``non-physical'' of degrees of freedom.

Thus, the Hamiltonian constraints are in fact a tool to define a probability measure for a manifold with a non-trivial topology (a principal fibre bundle for the gauge group)~\cite{gaugewhy}~\footnote{Despite that the gauge potentials can only be reconstructed up to one specific arbitrary point in space-time, the probability measure for a field with support only in that point is known because it is a probability measure for a finite-dimensional space.}, because a phase-space of gauge fields defined \emph{globally} on a 4-dimensional space-time (i.e. a fibre bundle with a trivial topology, when the base space is the Minkowski space-time) produces well-defined expectation functionals for the gauge-invariant operators acting on a fibre bundle with a non-trivial topology~\cite{gaugewhy}. 
On the other hand, setting non-abelian gauge generators to zero in the wave-function would require to solve a non-linear partial differential equation with no obvious solution~\cite{gaussYM,integralYM,globalYM,dressYM}~\footnote{Note that for the case of abelian gauge theories, we could use the Hodge decomposition~\cite{Ivancevic:2008dc} of the conjugate momentum $V_\mu=\partial_\mu \phi+V_\mu'$ of the potential $A_\mu$, where $\partial^\mu V_\mu'=0$ and we set $\phi=\Delta \rho$ whenever $\phi$ appears on the left or right extreme of an operator. Then the operator $\phi$ never acts on the cyclic vector generating the Hilbert space, so that it becomes irrelevant how would $\phi$ evaluate on the cyclic vector.}.

Note that it is crucial that the algebra of gauge-invariant operators is commutative. While this is not possible in the canonical quantization, it is possible with the quantization due to time-evolution as we will see later in this article. Note also that since only gauge-invariant operators are allowed, we must distinguish between the concrete manifold appearing in the phase-space and the family of manifolds (obtained from the concrete manifold through different choices of transition maps between local charts) to which the expectation values correspond. 

The gauge symmetry is different from anomalies.
An anomaly is a failure of a symmetry of the wave-function to be restored in the limit in which a symmetry-breaking parameter (usually introduced due to the mathematical consistency of the theory) goes to zero. We only consider symmetries of the Hamiltonian as candidate symmetries of the wave-function, since only these are respected by the time-evolution.

On the other hand, the constraints (which generate the gauge symmetry)
cannot modify the wave-functions of the Hilbert space.
Since in the case of a gauge symmetry there is no way to introduce a
symmetry-breaking parameter, we can never observe an anomaly.

\section{Lorentz covariance}
\label{sec:lorentz}

Concerning the relation between the Hamiltonian formalism and special relativity, there are two kind of questions we can ask: 1) is the Hamiltonian formalism compatible with Lorentz symmetry? 2) based on the space-time ``philosophy'', why should the time-evolution play a distinguished role in the Hamiltonian formalism?
The first question is technical, while the second question is conceptual. We do not have an answer to the second question, which is expected given the difficulties with Lorentz-symmetry of other approaches to quantization~\cite{segalvacuum}. But in this section we will answer explicitly and positively to the first question. In short, the fact that time evolution plays a special role allows us to use only Poincare representations with positive squared mass. Considering Poincare representations with positive squared mass is self-consistent and it is in no way in conflict with Lorentz symmetry.

A complete physical system is a free system. If we neglect gravity,
the wave-function associated to the free system is
a unitary representation of the Poincare group, regardless of the interactions
occurring within the free system~\cite{wigner}.

When the Hilbert space is the direct sum of irreducible representations of a symmetry group, then these representations will be defined by numbers (e.g. mass and spin) which are invariant under the symmetry group. Thus there will be a set of operators whose diagonal form corresponds to those invariants, we will call them Casimir operators.
When the symmetry group is abelian and continuous (e.g. translation in time), then the  generator of the group (e.g. the Hamiltonian) is a Casimir operator, and the invariant numbers defining the representations are called the constants of motion.
Certainly, when we move from non-relativistic quantum mechanics and consider instead the Poincare group, then the Hamiltonian is no longer a Casimir operator and the notion of constants of motion needs to be reviewed.
Nevertheless, the Casimir operators can be chosen arbitrarily (just like the Hamiltonian in non-relativistic quantum mechanics). 

For a positive squared mass, the spin and the sign of the Energy are also Poincare invariants.
The sign of the Energy times the modulus of the mass is the center-of-mass Energy, while the spin
is the center-of-mass angular momentum. Thus, the Casimir operator whose eigenvalues are the center-of-mass Energy may have negative eigenvalues and it will be the analogous operator to the Hamiltonian of the non-relativistic formalism. As will be seen in Section~\ref{sec:time}, such operator has the formal form of the Hamiltonian action (i.e. it is the difference between the generator of translations in time and the Hamiltonian operator). The Casimir operators necessarily commute with the momentum operator and thus they do not change the 3-momentum eigenstate. Thus we can solve the problem in a basis where the 
    3-momentum operator is diagonal.

In such a basis, the translations in space-time can be written as 
$T(x)\Psi(\gamma\vec{v})=e^{i M \tau(\gamma\vec{v}, x)}\Psi(\gamma\vec{v})$, where 
$\tau(\gamma\vec{v}, x)=\gamma x_0-\gamma \vec{v}\cdot \vec{x}$. 
Note that $\gamma=\sqrt{1+\gamma^2\vec{v}^2}$ is a function of $\gamma\vec{v}$. Thus in a basis where the 3-momentum is diagonal, the translations in space-time have the same structure as the time-evolution in non-relativistic space-time, with $M$ playing the role of the non-relativistic Hamiltonian and the numerical factor $\tau(\gamma\vec{v}, x)$ playing the role of the time (it is indeed the proper-time).

For each 3-momentum eigenstate, there is a corresponding inertial referential where the 3-momentum is null, i.e. the referential of the center-of-mass. In such referential, the modulus of the energy is the invariant mass, the signal of the energy is also a Lorentz invariant and the angular momentum is the spin.
Thus, the eigenvalues of the Hamiltonian and angular momentum operator in the center-of-mass define three Lorentz invariants which define the Poincare representation completely.


Despite we do not know a priori the
diagonal form of the Hamiltonian, we know that it is either continuous
or discrete in the neighborhood of the eigenvalue 0
(in the referential of the center-of-mass). 
If it is continuous then the zero energy has null measure. If it is discrete,
we can modify the Hamiltonian adding an appropriate constant such that the
zero energy is not one of the eigenvalues. Note that this is only possible
in a complex Hilbert space and this is equivalent to adding to the system a
free massive particle with null 3-momentum relative to the  system. In any
case, we can assume without loss of generality that our system is a quantum 
superposition of massive free systems with null 3-momentum. Then, the Lorentz
transformations become known and are given by the Wigner irreducible massive
representations of the Poincare group~\cite{wigner}. If the Hamiltonian is
bounded from below then the vacuum state is not Lorentz invariant, as it was
already suggested~\cite{segalvacuum}.

In the center-of-mass, the relevant group is not the Poincare
group, but the little group of spatial rotations and the translation
in time~\cite{wigner}. Thus the spatial and time coordinates of
space-time, become separated. The fields are no longer representations
of the Lorentz group, but only of the rotation group and the canonical
commutation relations are not in conflict with the little group
of spatial rotations. 

Note that we use Wigner's convention for the definition of the 3-momentum of the free complete system: it is the eigenvalue of the generator of the translations in space for the complete system (i.e. all fields defining the phase-space are translated in space). Thus in the center-of-mass, the algebra of operators has a constraint imposing that the operators are translation invariant. 

As it was discussed in Section~\ref{sec:constraints}, the cyclic vector defining the Hilbert space needs not be translation-invariant (in the center-of-mass), just the operators need to be translation-invariant in the center-of-mass.
This gives us a big freedom to choose  the cyclic vector defining the Hilbert space (which is related with the initial state of the system). 

We assume that the translations in space of the complete system
conserve the Hamiltonian and the constraints equations, such that
setting the total 3-momentum to zero in no way conflicts with the
constrained Hamiltonian system. Nevertheless, the restriction that there is a referential where the total 3-momentum is
null, excludes the free complete system from traveling at
the speed of light (e.g. a photon with non-null energy). Then the dynamics determined by the Hamiltonian becomes linked with the time coordinate (for a photon this would not be the case~\cite{diracdynamics}).

Therefore and unlike what it is often claimed in the literature, it is false that (canonical) quantization is incompatible with Lorentz covariance. Note that the phenomenologically successful (but ill defined) path integral formalism  based on the
Lagrangian is in fact equivalent to a path integral based on the
Hamiltonian~\cite{GrosseKnetter:1993ht}. In our formalism, the only restriction is that we need to consider representations with positive squared mass, then the dynamics determined by the Hamiltonian becomes linked with the time coordinate~\cite{diracdynamics}. The question why only positive squared masses are relevant is a reformulation of question 2) which will be left open in this paper. 
Similar assumptions concerning the energy-momentum of the full system are also done in the K\"{a}ll\'{e}n-Lehmann representation of a non-perturbative two-point correlation function, where it is assumed that the eigenvalues of the 3-momentum squared are not larger than those of the squared energy~\cite{Itzykson:1980rh}[p203]. 


The special role of the little group of rotations and the time evolution in our definition of a (special relativistic) Quantum Field Theory may seem to be a step back in the road towards a general relativistic quantum theory. In section~\ref{sec:gravity} we show that this is not the case, in fact our rigorous definition of a Quantum Yang-Mills theory marks \emph{``a turning point in the mathematical understanding of quantum field theory, with a chance of opening new horizons for its applications''} in quantum gravity, as requested in reference~\cite{prize}. 

We also make a comment on relativistic causality: the fact that we are considering only Poincare representations with non-negative squared mass leaves us in a good position to guarantee relativistic causality. However, we are working only in the 3-momentum space of the (free) complete quantum system. In order to study relativistic causality we need to make a unitary transformation to the 3-coordinate space, thus we need to define a position operator for a free quantum system. Defining a position operator is beyond the scope of this article, but it is done in another article~\cite{realpoincare}.



\section{Statistical Source Field Theory}
\label{sec:source}

The method of quantization described in Section~\ref{sec:time} is
inspired by the Source formalism of Schwinger~\cite{schwinger1989particles} which is itself both an
alternative to and inspired by the Feynman's path integral, where
time-ordering~\cite{timeordering,johnson2015feynman} plays a key role.

In this Section and in Section~\ref{sec:time} we will consider fields defined in a one dimensional time, 
neglecting the space dimensions. The extension of the results of this section to fields defined in four 
dimensional space-time is straightforward, since time and space are separated in the Hamiltonian formalism 
(see Section~\ref{sec:lorentz}).

We here use the term field meaning a function of a parameter $t$ which we also call timepiece. The timepiece has a close relation with the time, the main difference is that our fields are part of the phase-space of the theory: the state of the system is given by the functions of the timepiece $t$. Thus the phase-space has similarities to the space of trajectories in time of the Lagrangian formalism. Then, the
time-evolution will modify the state of the system as a function of the time parameter $\tau$.

Therefore, our fields are best described as source fields and we are
dealing with a statistical source field theory. Using a wave-function,
we can parametrize the probability distribution for a source field in
time. The linear space generated by all wave-functions is a
Fock space~\cite{rkhs}. The Fock space has the
properties of a continuous tensor product of Fock-spaces
corresponding to fields defined in infinitesimal time-intervals,
i.e. $\varphi(t)dt$. The time-evolution will not only advance the
time-intervals forward, but it will modify the wave functions
corresponding to each time interval accordingly to an Hamiltonian
which plays here the role of the connection in a covariant derivative. With abuse of
language, we can describe the situation as a continuous
tensor product of initial-value (i.e. Cauchy) problems, instead of
just one initial value problem as in standard Quantum Mechanics.

We have the self-adjoint position $x(t)$ and momentum $p(t)$
operators, verifying the Weyl relations.
\begin{align}
e^{i\int dt f(t) x(t)}e^{i\int ds g(s) p(s)}=
e^{-i \int dt f(t)g(t)} e^{i\int ds g(s) p(s)}e^{i\int dt f(t) x(t)}
\end{align}
where $f,g$ are real functions.

The Stone-von Neumann theorem implies that the Weyl relations uniquely
define the unitary operators $e^{i\int dt f(t) x(t)}$ and
$e^{i\int ds g(s) p(s)}$ up to a unitary transformation.

Thus, we can assume without loss of generality that the momentum and
position operator satisfy the canonical commutation relations:
\begin{align}
  [p(t), x(\tau)]=i\delta(t-\tau)
\end{align}
We can define a unitary translation operator as
\begin{align}
T(\tau) e^{i\int dt f(t) x(t)} T^\dagger(\tau)=e^{i\int dt f(t) x(t+\tau)}
\end{align}
and acting on the momentum operator in an analogous way.
We can express
\begin{align}
T(\tau)=e^{i \frac{\tau}{2} \int dt p(t)\partial_t x(t)-x(t)\partial_t p(t)}
\end{align}




The parameter $t$ from the phase-space and the parameter $\tau$ from the time-evolution are
deeply related and thus we call both parameters time, although they play different roles in our framework in the cases where the time-evolution $U$ has a non-trivial gauge symmetry.


If we consider the Hamiltonian $H$ and time-evolution $U$ defined as:
\begin{align}
H&=\int dt\ (p^2(t)+V(x(t),t))\\
U(\tau)&=T(\tau) e^{i \frac{\tau}{2} H}
\end{align}
Where $V(x(t),t)$ is a potential dependent on
the position operator and possibly also time-dependent.

We will use now the Trotter exponential product approximation~\cite{simon2005functional,Hatano:2005gh}, verifying for small $\epsilon$ and $A$, $B$, $A+B$ self-adjoint:
\begin{align}
e^{i\epsilon A}e^{i\epsilon B}=e^{i\epsilon
  (A+B)-\frac{\epsilon^2}{2}[A,B]+i\mathcal{O}(\epsilon^3)}
\end{align}
This is a good approximation since it works for unbounded self-adjoint operators $A,B$.

Then the time evolution is:
\begin{align}
U(\tau)&=e^{i \int_0^\tau dz \int dt\ p^2(t)+V(x(t),t-z) } T(\tau)
\end{align}
Where the exponential above stands for the time-ordered (with parameter $z$) product. Thus the Fock-space parametrization of a statistical field theory allows us to implement the concept of
time-ordering~\cite{timeordering,johnson2015feynman} consistently.

If we relax the mathematical rigor for a moment and imagine a source field completely localized
in one instant of time $t$, then the time-evolution (with time $\tau$) of that source field could
be described as a physical field function of time $\tau+t$ with initial conditions defined at time
$t$. 

In fact, a statistical source field theory can be seen as the solution of a particular Schrödinger equation. In this particular case, the Hamiltonian is given by the generator of the time-evolution $U$. Such generator does not depend on the parameter $\tau$ playing the role of the time in the Schrödinger equation. Note that the Hamiltonian does depend on the parameter $t$ but such parameter is part of the Fock-space which is a particular case of the Hilbert space appearing in the Schrödinger equation. 

For consistency with General Relativity (see Section~\ref{sec:gravity}), we also impose a constraint for the algebra of operators to be translation invariant in the time $t$. As it was discussed in Section~\ref{sec:constraints}, the cyclic vector defining the Hilbert space needs not be translation-invariant, just the operators need to be translation-invariant.
This gives us a big freedom to choose  the cyclic vector defining the Hilbert space (which is related with the initial state of the system).  

Note that for self-adjoint operators with null-expectation value, the contributions coming from the times $t$ corresponding to vaccuum expectation values are null. We assume that the translations in time $t$ of the complete system conserve the Hamiltonian and the constraints equations, such that setting the generator of translations to zero in no way conflicts with the constrained Hamiltonian system.

The main advantages of this more general formalism will be discussed in the next sections.



\section{Quantization due to time evolution}
\label{sec:time}

We introduce now the procedure of quantization due to unitary time
evolution.  Time evolution transforms a sequence of time-ordered
operators~\cite{timeordering} (which commute algebraically but the
time-ordering is non-commutative) into a sequence of 
(algebraically) non-commuting operators acting on a single slice of time of the wave-function. 
For a class of time evolutions (of the type of non-relativistic Quantum Mechanics), 
the canonical commutation relation of position and  momentum are reproduced 
(strictly speaking, it is the Weyl relation that is reproduced, 
i.e. the exponenciated version of the canonical commutation relation).

We use again the Trotter exponential product
approximation~\cite{Hatano:2005gh}, verifying for small $\epsilon$:
\begin{align}
e^{i\epsilon A}e^{i\epsilon B}=
e^{i\epsilon (A+B)-\frac{\epsilon^2}{2}[A,B]+i\mathcal{O}(\epsilon^3)}
\end{align}
This is a good approximation since it works for unbounded
self-adjoint operators $A,B$.

Let now $\epsilon=\frac{1}{n}$ with $n$ arbitrarily large. Then,
\begin{align}
e^{i\epsilon A}e^{iB}e^{-i\epsilon A}=(e^{i\epsilon A}e^{i\epsilon
  B}e^{-i\epsilon
  A})^n=e^{iB-\epsilon[A,B]+i\mathcal{O}(\epsilon^2)}
\end{align}
Therefore, for small enough $\epsilon$
\begin{align}
&e^{i\epsilon \int d\tau p^2(\tau)}e^{i\int dt f(t) x(t)}e^{-i\epsilon\int d\tau p^2(\tau)}
  =e^{i\int dt f(t) (x(t)+\epsilon p(t))}\\
&T(\epsilon) e^{i\epsilon \int d\tau p^2(\tau)}e^{i\int dt f(t) x(t-\epsilon)}
    e^{-i\epsilon\int d\tau p^2(\tau)}T^\dagger(\epsilon)=
    e^{i\int dt f(t) (x(t)+\epsilon p(t))} 
\end{align}

Now we need a definition of covariant derivative in time of the
position operator $x$, consistent with the fact that only
$F(a)=U(a)e^{i\int dt f(t) x(t-a)}U^\dagger (a)$ (where $U(a)$ is the time-evolution) is bounded while $x$ is unbounded. If we would be dealing with a commutative algebra, then the natural definition would be:
\begin{align}
\lim_{\epsilon \to 0} F^{\frac{1}{\epsilon}}(0)(F^{\frac{1}{\epsilon}}(\epsilon))^{\dagger}
\end{align}
For a trivial parallel transport, we would get as required:

\begin{align}
\lim_{\epsilon \to 0} F^{\frac{1}{\epsilon}}(0)(F^{\frac{1}{\epsilon}}(\epsilon))^{\dagger}=\lim_{\epsilon \to 0}e^{i\int dt f(t) \frac{x(t)-x(t-\epsilon)}{\epsilon}}
\end{align}

But since we are dealing with a non-commutative algebra, we need to use the
Trotter exponential product approximation formula, to define the
exponential version of the covariant derivative:

\begin{align}
&\lim_{\epsilon \to 0}\lim_{n \to \infty}(F^{\frac{1}{n\epsilon}}(0)(F^{\frac{1}{n\epsilon}}(\epsilon))^{\dagger})^n
\end{align}

And so for the parallel transport
$U(\epsilon)=T(\epsilon) e^{i\epsilon \int d\tau
  p^2(\tau)+V(x(\tau))}$ where $V(x)$ is a potential only dependent on
the position operator, the exponential version of the covariant
derivative of the position operator $x$ is:
\begin{align}
&\lim_{\epsilon \to 0}\lim_{n \to \infty}(F^{\frac{1}{n\epsilon}}(0)(F^{\frac{1}{n\epsilon}}(\epsilon))^{\dagger})^n=\\
&\lim_{\epsilon \to 0}\lim_{n \to \infty}=(e^{i\frac{1}{n\epsilon}\int dt f(t) x(t)} U(\epsilon)
                e^{-i\frac{1}{n \epsilon}\int dt f(t) x(t-\epsilon)} U^\dagger (\epsilon))^n=
    e^{i\int dt f(t) p(t)} 
\end{align}

The result is the exponential of the momentum operator, 
which verifies the Weyl relations with respect to the exponential of the position operator. 
With some abuse of language, we can say that for this type of time-evolution 
(and thus for this type of Hamiltonian $p^2+V(x,t)$, which is most common in non-relativistic Quantum Mechanics), the covariant derivative of the position operator is the momentum operator.
Thus the quantization (i.e. the Weyl relations) may appear in a statistical field theory due to a
particular non-deterministic time-evolution.

One major advantage of the quantization due to time-evolution is that it applies not only to the variables of
position and momentum which have an obvious correspondence in classical statistical mechanics, 
but it also applies to variables without an obvious correspondence in classical statistical mechanics such as 
fermions and ghosts (we just apply the time-evolution to the variables) 
and also in the presence of gauge symmetries as we will see in the remaining of this work.

\section{Gauge-variant gaussian measure and the phase diagram}
\label{sec:gaussian}

In the previous Section~\ref{sec:time}, we showed that the quantization due to time evolution works well for time evolutions of the kind of non-relativistic Quantum Mechanics. The question now is whether we can extend our results in a rigorous way to a Quantum Yang-Mills theory. However unlike in non-relativistic Quantum Mechanics, there is no rigorous definition of what is a Quantum Yang-Mills theory---since there is no theory to compare our results with, then our approach in the relativistic case is different than in the non-relativistic case. Our goal is to build a self-consistent rigorous theory which after some approximations (e.g. perturbative expansion, or a ultra-violet cutoff) reproduces the successful predictions of the Standard Model of Particle Physics. 

The Feynman's path-integral produces the notion of a gauge-invariant vacuum state, when the canonical formalism is ``derived'' from the path-integral formalism. However, the Feynman's path integral assumes the existence of a translation-invariant $\sigma$-finite (i.e. Lebesgue like) measure which is therefore gauge-invariant. Yet,
it is proved that in rigor such infinite-dimensional
Lebesgue measure cannot exist. As a consequence, the notion of a gauge-invariant vacuum state is inconsistent. On the other hand, a gauge-variant probability measure for an infinite-dimensional phase-space is mathematically consistent (e.g. a gaussian measure). 


As it was discussed in Section~\ref{sec:constraints}, some symmetries of the algebra of operators cannot be symmetries of the cyclic state defining the Hilbert space. 
In the case of gauge symmetry, the gauge potentials can be fully reconstructed from the algebra of gauge-invariant operators~\cite{wilsonloops}. Moreover, the Fock space (defined on a 4-dimensional space-time) produces well-defined expectation functionals for the gauge-invariant operators~\cite{gaugewhy}. The expectation-values of the gauge-invariant operators fully define the statistical gauge field theory (since the gauge potentials can be fully reconstructed~\cite{wilsonloops}), thus the gauge-variant operators can be neglected. Of course, gauge-variant operators can act on the Fock-space, but the link between these operators and the underlying manifold of gauge potentials is destroyed since the expectation-value is not gauge-invariant.

Since only (fully) gauge-invariant operators are allowed and the wave-functions necessarily break the gauge-symmetry, in scattering theory we always need to work in the in-in formalism. Of course, we can use the more common in-out formalism in intermediate steps. Explicitly, any bounded normal operator can be expressed in diagonal form using projection-valued measures. These projections are built using a basis of the Hilbert space, which can be expressed using gauge-invariant operators acting on the cyclic state (initial state). Thus, the complex amplitudes of the in-out formalism can be expressed as expectation values of the in-in formalism. The amplitudes are complex to allow that a constant can be added to the Hamiltonian without observable consequences, which is crucial to ensure the Lorentz covariance of the theory as it was discussed in Section~\ref{sec:lorentz}.

Thus the gauge-variant initial (cyclic) states are perfectly
fine, even at the non-perturbative level. In the canonical formalism the same gauge-variant initial states are allowed due to the adiabatic approximation and the Gell-Mann and Low's theorem~\cite{2007JMP....48e2113M,Brouder:2008xj}, although this issue is more subtle than in our formalism~\cite{Marzlin:2004zz}.

This does not imply that the mean-field approximation (that is, the usual choice of exact wave-function around which perturbative corrections are applied) always works, but the non-perturbative problems of the mean-field approximation are not exclusive to the gauge-symmetry (e.g. the mean-field approximation also breaks global symmetries in the two-Higgs-doublet model). So it can happen that within our (non-perturbative) mathematical definition of Quantum Gauge Field Theory, some exact predictions differ substantially from the corresponding perturbative approximation, being the predictions involving non-null non-abelian global charges obvious candidates for this difference to show up~\cite{gaussYM,integralYM,globalYM,dressYM}. However, the non-perturbative validity of our definition also shows that there is no fundamental reason why all non-abelian global charges must be null as claimed recently~\cite{Maas:2017xzh}; this is welcome since we know that (probably) in quantum gravity there is a non-abelian gauge symmetry (torsion) whose corresponding global charge is the total spin of the quantum system which must be allowed to be non-null otherwise quantum gravity would be incompatible with the experimental data.

The global charge operator is different from a linear combination of gauge generators which are constrained to be zero (due to surface terms at infinity). In the in-in formalism, we can characterize an initial state with a non-null global charge using only gauge-invariant operators (e.g. the Casimir operators of the algebra generated by the global charge of an abelian or non-abelian gauge-theory, are gauge-invariant operators). 

\section{Local operators and the momentum constraint}
\label{sec:local}

The momentum constraint also generates a gauge symmetry, once we consider a spectral measure where the fields are functions of space and time. Then the momentum constraint always modifies the spectral measure and so we have a complete non-comprehensive gauge-fixing.

Since all operators must be invariant under a translation in space, how can we define local operators? In rigor we can't, but we can define operators which behave effectively as local operators. We just need to add  3-dimensional position quantum numbers to the Hilbert space. These quantum numbers do not appear in the Hamiltonian, so they are invariant under translation in time. Then the operator $\int dP_{\vec{x}} l(\vec{x})$ is translation invariant, where $dP_{\vec{x}}$ is the spectral measure corresponding to the new position quantum numbers and $l(\vec{x})$ is the operator we want to evaluate locally. The translation-invariant operator behaves effectively as a local operator when the wave-function is concentrated around one point in the new 3-dimensional space. Note that then the wave-function is not translation invariant which is fine, as long as the operators are translation invariant.

\section{Relation with the BRST formalism}
\label{sec:brst}

The results of reference~\cite{Burnel:2008zz}, are consistent with our formalism:\\
1) there is no fundamental reason why BRST-like gauge-fixing should be Lorentz invariant;\\
2) but it is crucial that the BRST-like gauge-fixing term involves the time derivative of the fields which have an arbitrary time-evolution, otherwise the theory becomes non-local and the perturbative expansion becomes inconsistent (as it happens in the Coulomb gauge, see also~\cite{Rothe:2010dzf,photonlocalization});\\
3) for technical reasons that (apparently) are not related with Lorentz covariance, the $R_\xi$ gauges are better suited for perturbation theory than any other gauge (Lorentz covariant or non-covariant).

The result 1) is consistent with our formalism where the fields are representations of the little group of rotations and not of the Lorentz group. In our formalism, we do not need the BRST-like gauge-fixing to define the theory, however to go from a phase-space defined in space-time to a phase-space defined in space only we need to  fix the time-evolution of all fields. As we will see in this Section, this is done using a BRST-like gauge fixing in agreement with result 2).
Finally, there does not seem to exist an obvious reason for result 3), but in any case our formalism is not incompatible with the result 3).

We are working from the start with a self-consistent statistical field theory.

The BRST charge is useful for a non-commutative algebra, because when we multiply the right and left ideal the result would not be an ideal if the BRST charge would not square to 0. The alternatives would be to use (standard) gauge-fixing, which suffers from nonlocality in general and suffers from the Gribov problem in non-abelian gauge theories; or to work only with a commutative sub-algebra of the algebra of operators satisfying the constraints, which is challenging for a non-deterministic time-evolution.

The algebra of operators is then enlarged from gauge-invariant operators to BRST-invariant (not necessarily gauge-invariant) operators. These BRST-invariant operators are divided into equivalence classes. The BRST cohomology maps (in a bijective way) each equivalence class with a corresponding gauge-invariant operator satisfying the constraints. When performing algebraic manipulations in a gauge-invariant operator,
 we can convert the gauge-invariant operator into BRST-invariant, then do the manipulations in the space of BRST-invariant operators and then convert the resulting BRST-invariant operator again into a gauge-invariant operator.

Note that there are two possible inner-products: one degenerate where the ghost fields are self-adjoint and another non-degenerate where the ghost fields are not self-adjoint and behave like standard fermioninc creation and annihilation operators~\cite{VanHolten:2001nj}[Sec.2.7]. We use the non-degenerate inner-product, so that we are always working within an Hilbert space formalism. We can do this, because the vacuum state in our formalism is not Lorentz invariant if it exists and the existence of such vacuum state is not mandatory for our formalism. This would not possible in the covariant operator formalism~\cite{cof} where the vaccuum is Lorentz invariant. Since our vacuum breaks Lorentz invariance, then the spin-statistics theorem does not hold and the ghosts are as consistent as a Schrödinger field. Moreover, since the ghosts only appear in intermediate stepts and the ghosts never appear in the definition of expectation values of observables (neither in the operators nor in the wave-function), then the complete system when treated as a free particle respects the spin-statistics theorem.



Crucially, the BRST cohomology merely simplifies the expression defining a gauge-invariant operator into another equivalent expression, it does \emph{not} affect the gauge-variant wave-function and therefore the Gribov problem does not arise.

 

There is a well-known subtlety with the BRST cohomology that we need to address: the BRST cohomology is itself gauge-invariant and mathematically well-defined, but it is merely a dispensable auxiliary step in a calculation performed in the context of a quantum formalism. If the quantum formalism is mathematically inconsistent, if the formulation of the calculation crucially depends on the BRST-invariant algebra (not our case, but it is the case of the path integral), surprises are possible.
In particular, if all the details of the calculation are only known for the BRST-invariant algebra and not before for the gauge-invariant operators (not our case, but it is the case of the quantum BRST formalism), then the gauge-invariance of the BRST cohomology does not imply that the calculation would be the same in all gauges or that the quantum formalism is logically consistent (and in fact it is not due to the Gribov problem). This is discussed in reference~\cite{brst_gaugeinvariance}:

\epigraph{``Being gauge invariant, the BFV-PI necessarily reduces to an integral over modular space,
irrespective of the gauge fixing choice. Nevertheless, which domain and integration measure
over modular space are thereby induced are function of the choice of gauge fixing conditions.
The BFV-PI is not totally independent of the choice of gauge fixing fermion $\Psi$.''}

In the (our) case of the quantization due to time-evolution, the quantum formalism is mathematically well-defined and all details of the calculations are known, regardless of whether we apply the BRST cohomology or not. Since we use the BRST cohomology to merely simplify the expression defining a gauge-invariant operator into another equivalent expression, the Gribov problem does not affect us.

\section{Renormalization, the mass gap and the Millennium prize}
\label{sec:rg}

We couldn't propose a mathematical definition of a Quantum gauge theory without discussing renormalization and the Millennium prize (Clay Mathematics Institute)~\cite{prize}, which are related.

The Millennium problem \emph{``Yang-Mills and Mass Gap''} defined by the Clay Mathematics 
Institute~\cite{prize}, consists essentially in defining a Quantum Yang-Mills theory in a mathematically rigorous and useful way. By useful, we mean that the definition must mark \emph{``a turning point in the mathematical understanding of quantum field theory, with a chance of opening new horizons for its applications''} (as stated in reference~\cite{prize}).

Note that there is a monetary prize associated to the problem. If the requirement would only be that the definition would be mathematically rigorous, then (as so often happens in mathematics) the definition could have no implications to theoretical physics. No one doubts that many mathematical entities resembling a quantum Yang-Mills theory exist, the relevance of finding ``the'' definition lies in the hope that it will lead to progress not only in mathematics but also in physics.

On the other hand, ``useful'' is not an objective criterium. Since the authors of reference~\cite{prize} believed that any useful definition of a Quantum Yang-Mills theory would necessarily establish the existence of a mass gap, then they added the requirement that the definition of Quantum Yang-Mills must be such that it establishes to the existence of a mass gap, that is, that the spectrum of the Hamiltonian is discrete close to the ground state.

The reason that the authors of reference~\cite{prize} believed so, is that the existence of a mass gap implies clustering which is a locality property required (from a technical point-of-view) to extrapolate many results obtained in simplified theories to a realistic Quantum Field Theory. Admitting the possibility that a useful definition of Quantum Yang-Mills theory would not prove the existence of a mass gap, would be admitting that a significant part of the scientific work made by the mathematical and theoretical physics community in the last 50 years is clearly speculative.

Note that from a physics perspective, the existence of a mass gap in any not too simple theory is not surprising at all. Already for Quantum Electrodynamics in a 3 dimensional space-time (in the lattice approximation at least~\cite{qedpolyakov}) there is a mass gap and confinement even at weak coupling.

With respect to the mass gap, our definition of a Quantum Yang-Mills theory is better than the type of definition proposed by the authors of reference~\cite{prize}. Since the Hamiltonian of our theory is well defined, whether or not there is a mass gap is entirely determined by the Hamiltonian and the initial (cyclic) state. As it was mentioned above, it would not be surprising if a specific theory has a mass gap because the mass gap is a non-perturbative feature. For the same reason, it is extremely difficult to find a large class of theories where there is necessarily a mass gap.  Our definition of a Quantum Yang-Mills theory is valid whether there is a mass gap or not, therefore there is no need (as required by the type of definition proposed by the authors of reference~\cite{prize}) to find a large class of theories where there is necessarily a mass gap which would be extremely difficult. 

The above discussion also has implications for the confinement mechanism in the context of Quantum Chromodynamics. There is no need for any special mechanism to ``hide'' the gauge fields at low energies. The common fact that the bound states are in this theory the states of lowest energy is entirely determined by the Hamiltonian of the theory and suffices to ``hide'' the massless gauge fields, as it happens for Quantum Electrodynamics in a 3 dimensional space-time for any value of the gauge coupling~\cite{qedpolyakov}. 

The propagation of an excitation (a photon, a gluon) in a medium is best described in an effective field theory framework~\cite{eft_quasiparticle}. Renormalization is an essential part of an effective field theory and it describes the quasi-(free) particle properties of the excitation (such as the effective mass), whether the associated perturbation theory needs to be regularized or not due to ultra-violet divergencies~\cite{weinberg}.

Concerning regularization, we certainly need to define an Hamiltonian with respect to an Hilbert space and for that reason we use the (symmetric) Weyl ordering~\cite{Fujii:2003ax, weylordering} due to the fact that it conserves the exponential of operators, unlike normal-ordering~\cite{normalordering}. This is an important property of the ordering, because we use often the Trotter product formula. Note that we also subtract the vacuum expectation value of the self-adjoint operators (such as the Hamiltonian, see Section~\ref{sec:source}), otherwise we may have divergences. 

But without perturbative expansions, the initial state needs not to be an eigen-function of the Hamiltonian (and usually it is not), so that we do not have to eliminate divergences due to tadpoles~\cite{Ellis:2015xwp}; any initial state is a good asymptotic state so we have no infrared divergences~\cite{finite};
and we do not have loops introduced by the perturbative expansion so no ultra-violet divergences~\cite{finite}.

That is, our theory needs no regularization. 
Subtracting the vacuum expectation value of the self-adjoint operators is not by itself regularization because there is no regularization parameter.
Note that a regularized theory is necessarily an effective field theory, where the renormalization plays a key role.

Our definition of Quantum Yang-Mills theory is thus not an effective field theory.
Thus the renormalization group in our theory cannot be distinguished from a regular background symmetry group:
that is, a symmetry group that acts not just on the fields but also on the parameters of the theory, leaving the observables invariant. The renormalization group plays no fundamental role in our definition of Quantum Yang-Mills theory. This is the ultimate reason why establishing the existence of a mass gap has nothing to do with the problem of defining the theory (at least in the way we do it). And this is a good property of our definition, since proving the existence of a mass gap and defining the theory simultaneously would be much more complex, if possible at all.

\section{Pure Yang-Mills theory}
\label{sec:ym}

The $SU(N)$ index $\{a,b,c\}$ allows us to cover also the electromagnetism by eliminating the 
$SU(N)$ index and thus setting the structure constants $f_{a b c}$ to zero.
We used the Cadabra software~\cite{Peeters2018,Peeters:2007wn} to confirm the calculations.
We follow reference~\cite{VanHolten:2001nj}:
\begin{align}
\{\psi_{a}(x),\psi^\dagger_{b}(y)\}=&\psi_{a}(x) \psi^\dagger_{b}(y)+\psi^\dagger_{b}(y)\psi_{a}(x)
=\delta_{a b}\delta^4(x-y)\\
[A_{\mu a}(x),\pi^{\nu}_{b}(y)]=&A_{\mu a}(x)\pi^{\nu}_{b}(y)-\pi^{\nu}_{b}(y) A_{\mu a}(x)=
i\delta^{\nu}_{\mu} \delta_{a b}\delta^4(x-y)\\
\Omega=&\int d^4x \left[ \pi^{\mu}_{a} \partial_{\mu} \psi^\dagger_{a}-\pi^{\mu}_{a} f_{a b c} A_{\mu b} \psi^\dagger_{c} -i \frac{1}{2}f_{a b c} \psi^\dagger_{a} \psi^\dagger_{b} \psi_{c}\right]
\end{align}
where $\Omega$ is the generator of the BRST symmetry and the structure constants $f_{a b c}$ are related with 
the $SU(N)$ generators $T_{a}$ by:
\begin{align}
[T_{a}, T_{b}]=i f_{a b c} T_{c}\\
\tr(T_a T_b)&=\frac{1}{2}\delta_{a b}
\end{align}

The covariant derivative is given by:
\begin{align}
D_{\mu}&=\partial_{\mu} -i g T_a A_{\mu a}\\
[D_\mu,D_\nu]&=-ig T_a F_{\mu\nu a}
\end{align}

The magnetic components of the gauge field strength tensor are given by: 
\begin{align}
B_{i a}&=\frac{1}{2}\epsilon_{i j k} (\partial_j A_{k a}-\partial_k A_{j a}-g f_{a b c} A_{j b} A_{k c})\\
B_{i}&=i\frac{1}{2 g}\epsilon_{i j k} [D_j, D_k]=B_{i a} T_a
\end{align}
Where the indices correspond to the spatial dimensions only, i.e. from 1 to 3. It verifies the Jacobi relation
\begin{align}
[D_i,B_{i}]=i\frac{1}{2 g}\epsilon_{i j k}[D_i,[D_j, D_k]]=0
\end{align}

We need to separate the ideal (gauge generator) from the gauge-invariant algebra. That is, not only the gauge-invariant algebra must commute with the ideal, but also the ideal cannot be included in the gauge-invariant algebra. This is guaranteed by (non-comprehensive) gauge-fixing as it was discussed in Section~\ref{sec:constraints}. For instance, in the case of the free Electromagnetic field, while the gauge generator commutes with the ideal (because the ideal is the gauge generator itself) it cannot be included in the gauge-invariant algebra; thus, the Electromagnetic field as a whole is not part of the gauge-invariant algebra. And that is a good thing, since then we always work with local field operators, unlike those of the Coulomb gauge.

The Hamiltonian for the Yang-Mills theory in our formalism has the same form as the classical Hamiltonian Action~\cite{Mukhanov:1994zn}:
\begin{align}
  H=\int d^4 x \left[\pi^i_a \partial_0 A_{i a}-A_{0 a} (D_i \pi^i)_{a}-\frac{1}{2}\pi^i_a\pi^i_a
  -\frac{1}{2} B_{i a} B_{i a}\right]
\end{align}

Due to the BRST cohomology, we can add to the Hamiltonian the term: 
\begin{align}
\{\Omega, i \psi_{a} A_{0 a}\}=A_{\mu a}\partial_0\pi^{\mu}_{a}-i\psi^\dagger_{a}\partial_0\psi_{a}
\end{align}

We obtain a BRST invariant Hamiltonian in the Weyl gauge:
\begin{align}
H_W=H+\{\Omega,\int d^4 x \psi_a A_{0 a} \}=\int d^4x \left[\psi^\dagger_{a} \partial_0 \psi_{a} 
+ \pi^\mu_{a} \partial_0 A_{\mu a}-\frac{1}{2}\pi^i_a \pi^i_a-\frac{1}{2} B_{i a} B_{i a}\right]
\end{align}

Since the Weyl gauge is an incomplete gauge-fixing, this new Hamiltonian $H_W$ is invariant under a remnant
gauge symmetry generated by $\theta_a (D_i \pi^i)_{a}$, where the function $\theta$ is constant
in time. The translation in time can now be factorized, allowing us to redefine a new Hamiltonian constant
in time:
\begin{align}
\int H_3(x^0)dx^0=&H_W-\int d^4x \left[c^\dagger_a \partial_0 c_a-\pi^\mu_a \partial_0 A_{\mu a}\right]
\end{align}
Where $H_3(x^0)$ is the new Hamiltonian constant in time and integrating 3D space only.

Considering only operators respecting time-ordering, we can move all operators to a single slice of time
and then the predictions coincide with the Dirac quantization of the Yang-Mills theory
(and also for the Electromagnetism).
We can now reapply the BRST cohomology for the remnant gauge symmetry (as discussed in
reference~\cite{McKeon:2014lwa}). The remnant gauge transformations are constant in time.

The time-evolution and the remnant gauge transformations of $A_{0 a}$ decouple from the time-evolution
and the remnant gauge transformations of $A_{i a}$ (respectively); $A_{0 a}$ commutes with the remnant gauge transformations. Thus, the gauge fields $A_{0 a}$ can be studied first, without taking into account the 
$A_{0 i}$ fields. We do this by integrating only the $A_{0 a}$ fields, while keeping the $A_{i a}$ fields as external variables to the theory and within the theory it is as if we would only consider operators which do not depend on $A_{i a}$. Then, we are left with the $A_{0 a}$ fields only, which commute with $H_3(x^0)$ and also with the remnant gauge transformations, then the corresponding gauge theory is that of a gauge-invariant field which is constant in time, which has an easy solution.

After integrating the $A_{0 a}$ fields, we obtain operators which depend on $A_{i a}$ only and do not depend on $A_{0 a}$. Note that in the literature (and in reference~\cite{henneaux1992quantization} in particular) it is usually claimed that the operators do not depend on $A_{0 a}$ because they were chosen not to depend on 
$A_{0 a}$ (due to an argument related to gauge-invariance as we will see below) and not because the fields 
$A_{0 a}$ were integrated first. This is not incompatible with our results because the remnant-gauge-invariant operators involving $A_{i a}$ only, are also fully gauge-invariant. Thus our framework is more general but it is compatible with reference~\cite{henneaux1992quantization} because we can also simply choose remnant-gauge-invariant operators involving $A_{i a}$ only, which will be automatically fully gauge-invariant even if we just ignore the $A_{0 a}$ fields.

Considering the remnant gauge-symmetry and only a single slice of time, we can then just follow
reference~\cite{henneaux1992quantization}[pp460] on how to choose the Lorentz gauge. The first step is to introduce a new field $\Phi$ and a new constraint $\Pi=0$ where $\Pi$ is the conjugate field of $\Phi$. Such a field is completely spurious because it allows a complete and comprehensive gauge-fixing, thus we can remove it from the gauge-invariant operators (as expected, otherwise we would have a different theory from the one we started before introducing the new field).

\section{Free electromagnetic field: an exact example}
\label{sec:em}

While the free electromagnetic field is an abelian gauge theory and thus it is somewhat simpler than a Yang-Mills theory (for instance, there is no Gribov ambiguity), it has the crucial advantage that the time-evolution has an exact solution, because the Hamiltonian is quadratic in the fields.

The extrapolation of the results of the previous section to the electromagnetism is straightforward since
the Yang-Mills theory is an obvious generalization of the electromagnetism. The contraction of the index is just an integration in a 4-dimensional space-time and we set the structure constants $f_{a b c}$ to zero.

Due to the complete non-comprehensive gauge-fixing, we always start from the magnetic components of the gauge field strength tensor $B_{i}$ (for local self-adjoint operators). Using a (remnant from the Weyl gauge) gauge-invariant unitary transformation (such as time-evolution) we can generate other operators, since the unitary transformation can be transferred from the operators to the wave-function without modifying the constraints. Thus, through time-evolution (in the Weyl gauge), we also have the local self-adjoint operator $\vec{\partial} \times \vec{\pi}$, i.e. the curl of the Electric Field. It is not possible to use just the Electric Field or its divergence as an operator acting on the Hilbert space, which is a sign of consistency because the divergence of the Electric Field is a constraint set to zero before the definition of the Hilbert space. This allows us to always work with local operators, which avoids the localization problems which arise when working with non-local transverse fields~\cite{photonlocalization}.

\section{A translation-invariant measure and classical field theory}
\label{sec:measure}

While there is a mathematically rigorous definition of classical field theory~\cite{cftmath}, so far the definition of a (classical) statistical field theory is tied to the definition of a quantum field theory, which involves an ultraviolet microscopic scale related to a lattice spacing necessary to regularize the ultraviolet divergencies of the field theory and to implement its renormalization~\cite{mussardosft}. We have shown in the previous sections that for Hamiltonians which are polynomial in the fields, our mathematical formalism based on the gaussian measure (Fock-space) is much more powerful and general than the peculiar notion of ``continuum'' through renormalization, starting from the fact that our formalism involves the continuum independently from renormalization and it can also be renormalized (to reproduce the standard perturbative calculations, for instance).

However, the Fock-space only allows for polynomial Hamiltonians. This excludes a rigorous definition of the classical statistical version of many classical field theories (such as General Relativity), since so far there is no reason why the Hamiltonian of a classical field theory should be polynomial in the fields, not to mention Quantum Gravity~\cite{Katanaev:2005xd}. This is unnacceptable: for most classical field theories, the definition of the corresponding classical statistical field theories should be straightforward. We always need a statistical theory, because the real-world measurements are never fully accurate.

This is an indication that an alternative to a gaussian measure which allows the definition of non-polynomial Hamiltonians should not be too hard to find. Indeed, the essential obstruction to an infinite-dimensional Lebesgue measure is its $\sigma$-finite property~\cite{baker1991lebesgue,baker2004lebesgue}. Once we drop the $\sigma$-finite property, several relatively simple candidates exist~\cite{baker1991lebesgue,baker2004lebesgue}.

Since the Fock-space allows us to define successfully the Standard Model of Particle Physics, we expect the alternative to a gaussian measure to be closely related to the Fock-space and to somehow coincide with the Fock-space for polynomial Hamiltonians. There is an important theorem about Fock-spaces, which is useful for our case~\cite{partitionfock,skeide,indicator,indicator2} stating that the closed linear span of exponential vectors:

\begin{align}
\biggl\{e(\chi_{[s_1,t_1]}+\cdot\cdot\cdot+\chi_{[s_n,t_n]}):
0 \leq s_l \leq t_1 \leq s_2 \leq t_2 \cdot\cdot\cdot s_n \leq t_n,\ n\geq 1\biggr\}
\end{align}
is $\Gamma(L^2(\mathbb{R}_+))$.

Since the Fock space $\Gamma(L^2(\mathbb{R}))$ is isomorphic to the tensor product of two $\Gamma(L^2(\mathbb{R}_+))$, then the theorem can be extended to $\Gamma(L^2(\mathbb{R}))$. This implies that the Fock-space on $\mathbb{R}^4$ can be generated by the tensor products of an (arbitrary but) finite number of Hilbert-spaces corresponding to different regions of $\mathbb{R}^4$, with the vaccuum state corresponding to the remaining region of $\mathbb{R}^4$. If we also attribute one Hilbert-space to the remaining region of $\mathbb{R}^4$ instead of the vaccuum state, we still generate an Hilbert-space (although one which is unitarily inequivalent to the Fock-space, instead the Fock-space is a subspace of this new Hilbert space): this is the space $L^2(S_m)$ where $S_m$ is the linear space of simple real functions: $\sum_{k=1}^m a_k 1_{A_k}$, where $a_k$ are real numbers and $1_{A_k}$ are indicator functions with $A_k \subset \mathbb{R}^n$. Note that any complex function $f$ appearing in $e^{a^+(f)}$ is $f\in L^2(\mathbb{R}^4)$ and thus it is the limit of a sequence of simple complex functions.
Since the coherent states are overcomplete, to each complex number appearing in the simple complex functions it corresponds one Hilbert space $L^2(\mathbb{R})$.
In this way we generate an effective Hilbert space of square-integrable functions (not necessarily polynomials) whose domain is a linear space of (other) real functions which are $L^2(\mathbb{R}^4)$ up to a constant function.

The requirement of real functions which are $L^2(\mathbb{R}^n)$ up to a constant is crucial, and it determines the field configurations which are allowed. For instance, only fields which converge in infinity are allowed (hence square-integrable...), but the value of the field at infinity is arbitrary (hence ...up to a constant) so that spontaneous symmetry breaking is possible  (see Section~\ref{sec:spontaneous}). Also, the fields form themselves an Hilbert space $L^2(\mathbb{R})$, so that the value of the field in a domain with null Lebesgue measure is ignored (this excludes Dirac deltas, for instance). 

The resulting measure is translation-invariant, but not $\sigma$-finite (as expected) because it is not the countable direct sum of finite measures.  Thanks to the close-relation to Fock-spaces, we know how to do calculations with such measure. Note that the linear space of fields has no gauge constraints just like the Fock-space, so there are no contradictions with Section~\ref{sec:gaussian}.

There is one crucial difference with respect to the Fock-space: the vacuum expectation value of an operator is replaced by the operator evaluated with a configuration of the fields constant (with the same value as at infinity, for consistency reasons). We define the field $\phi_k$ constant in the subset $A_k\subset \mathbb{R}^4$ with respect to the spectral measure  $d\phi_k$, while the canonical conjugate field is $i\frac{\partial}{\partial\phi_k}$. Note that the Stone-von Neumann theorem applies to each element of the basis of the Hilbert space---because it involves simple functions only---, so that there is no room for surprises when defining the fields and its canonical conjugates. We may now define Hamiltonians which are non-polynomial functions of the field $\phi_k$ with respect to the spectral measure  $d\phi_k$, just like it happens in Quantum Mechanics (e.g. the Hamiltonian for the Hydrogen atom). We can always drop the zero-point field Energy from the Hamiltonian~\cite{quantumvacuum}, thus the Hamiltonian can always be made to asymptotically vanish even when there is spontaneous symmetry breaking (see Section~\ref{sec:spontaneous}).

\section{Spontaneous symmetry breaking}
\label{sec:spontaneous}

There are several definitions of spontaneous symmetry breaking in the context of statistical mechanics, based on: a long-range order parameter which is the expectation value of a function $f(A)$ invariant under a group $G$ (e.g. the modulus $f(A)=|A|$) of an operator $A$ which is translation invariant and breaks $G$; or a conditional expectation value of some operator $A$ given some condition $C=0$ that breaks the symmetry; or a two-point correlation function with the points at an infinite distance from each other (related with boundary conditions)~\cite{yangising}.

In the translation-invariant measure discussed in Section~\ref{sec:measure}, the field configurations converge in infinity to an arbitrary constant value so that spontaneous symmetry breaking is possible.

As it was discussed in Section~\ref{sec:gaussian}, the wave-function necessarily breaks the full gauge symmetry. But it may or not break the global gauge symmetry (remnant of the full gauge symmetry). However as it was discussed in Section~\ref{sec:constraints}, there is no spontaneous symmetry breaking of the (full or global) gauge symmetry, since the expectation-values of the gauge-variant operators are null due to the constraints. Note that the gauge-invariant operators are not sensitive to the long-range correlations which could signal spontaneous symmetry breaking~\cite{Elitzur:1975im}.

Still the phase diagram of the theory may be sensitive to whether or not the field configuration at infinity conserves the global gauge symmetry. As it was discussed in Section~\ref{sec:constraints}, since only gauge-invariant operators are allowed, we must distinguish between the concrete manifold appearing in the phase-space and the family of manifolds (obtained from the concrete manifold through different choices of transition maps between local charts) to which the expectation values correspond. Thus, the concrete manifold appearing in the phase-space may feature spontaneous symmetry breaking of the global gauge symmetry and this may cause effects in the phase-diagram (such as the existence of a phase transition), despite that there is no spontaneous symmetry breaking of the (full or global) gauge symmetry; the Higgs mechanism is an example of such a case.


\section{Quantum gravity}
\label{sec:gravity}


The first step towards Quantum Gravity, is to choose which version of (classical) General Relativity we want to quantize and the corresponding definition of space-time.
For that purpose, it is useful to review the case of the Yang-Mills theory exposed in the previous sections: in a non-abelian gauge-theory, the Gribov ambiguity forces us to consider a phase-space formed by fields defined on not only space but also time. This is related to the fact that in a fibre bundle (the mathematical formulation of a classical gauge theory) the time cannot be factored out from the total space because the topology of the total space is not a product of the base-space (time) and the fibre-space, despite that the total space is \emph{locally} a product space. Thus, the Hamiltonian constraints are in fact a tool to define a probability measure for a manifold with a non-trivial topology (a principal fibre bundle for the gauge group) such that the only measurable functions are the gauge-invariant functions~\cite{gaugewhy}, because a phase-space of gauge fields defined \emph{globally} on a 4-dimensional space-time (i.e. a fibre bundle with a trivial topology, when the base space is the Minkowski space-time) produces well-defined expectation functionals for the gauge-invariant operators acting on a fibre bundle with a non-trivial topology~\cite{gaugewhy}.

Since only gauge-invariant operators are allowed, we must distinguish between the concrete manifold appearing in the phase-space and the family of manifolds (obtained from the concrete manifold through different choices of transition maps between local charts) to which the expectation values correspond. Regarding General Relativity,
the tetrad fields can be reconstructed from diffeomorphism- and local Lorentz-invariant operators~\cite{grgauge}, as in the reconstruction theorem of Yang-Mills theory~\cite{wilsonloops}. The concrete space-time has a spinor structure, otherwise it would be necessary to define spinor fields which transform non-linearly under the diffeomorphism group~\cite{Pitts:2011jv}. Also, a noncompact (4 dimensional) space-time admits a spinor structure if and only if it has a global field of orthonormal tetrads~\cite{geroch}; meaning that the concrete space-time is parallelizable~\cite{geroch} and so the Levi-Civita and the Weizenbock connections coexist~\cite{Youssef:2006sz} and General Relativity may be equivalently defined as Teleparallel Gravity~\cite{aldrovandi2012teleparallel}. Many well-known solutions of the Einstein equations do have a spinor structure~\cite{geroch}. However, if only diffeomorphism- and local Lorentz-invariant operators are allowed, then the expectation values would correspond to any space-time with a Lorentz metric as if we would have defined the tetrads only \emph{locally}.

In fact, the space-time in the Hamiltonian formalism is time-orientable since it requires a global time-like vector field giving the direction of the time-evolution~\cite{gerochglobal}; the space-time also requires a volume form (to define the Hamiltonian) which is an orientation~\cite{guillemin2010differential}; the space-time is also space-orientable, because it is orientable and time-orientable~\cite{gerochglobal}. Both the volume form and the global time-like vector field are invariant under local Lorentz transformations. 
Crucially, the diffeomorphism invariance of the Hamiltonian formalism remains intact.

We stress that a parallelizable space-time doesn't need to have a Cauchy surface because the derivatives corresponding to different tetrads do not need to commute, thus the parallelizable space-time doesn't need to be globally hyperbolic (i.e. topologically a product space of time and a 3-dimensional manifold)~\cite{dieckmann,Geroch:1970uw,geroch,geroch2}.

If we assume that there is also a global scalar field corresponding to time (which implies that the space-time has stable causality~\cite{gerochglobal}), then we can define a family of 3-dimensional achronal (but not necessarily space-like) hypersurfaces corresponding to constant times~\cite{gerochglobal}. However, the step from local existence of solutions to the Einstein equations to the existence of a maximal development is non-trivial, due to the diffeomorphism
invariance of the equations~\cite{ringstrom2009cauchy,Kiriushcheva:2008sf}. It turns out that given initial data (that is, one 3-dimensional hypersurface corresponding to a specific time), there is a unique maximal globally hyperbolic development~\cite{ringstrom2009cauchy}. Despite that the maximal development is only a subset of the total space-time, it is nevertheless a space-time in itself which can be defined by an Hamiltonian formalism. This justifies the success of the predictions of numerical relativity based on the ADM formalism~\cite{constraintalgebra,Deser:1976ay}) which assumes an orientable globally hyperbolic spacetime (that is, all of the 3-dimensional achronal hypersurfaces are space-like). 

Thus, there is no guarantee ``a priori'' that in Quantum Gravity the space-time is globally hyperbolic as it is assumed in the quantum ADM formalism. Moreover, the classical ADM formalism is a weakly hyperbolic system of partial differential equations and thus it is divergent and not well-posed~\cite{Kiriushcheva:2008sf}. Thus, there is no classical Hamiltonian formalism which is well-posed. Note that the ``Hamiltonian formalism'' proposed in reference~\cite{Kiriushcheva:2008sf} is admittedly only definined locally and thus it is not a solution to the Cauchy problem in General Relativity as a complete Hamiltonian formalism should be.


We also assume that the space-time is asymptotically flat, for the sake of simplicity when defining the probability measure (see Section~\ref{sec:measure}) although this assumption is not strictly required. 

The second step towards Quantum Gravity is to choose the mathematical tools needed to define the theory.
There is a widespread belief that the sequence of generalizations on the descriptions of space and time 1) Galilean invariance, 2) special relativity, 3) general relativity; which happened for deterministic theories should also happen for quantum theories~\cite{Isham:1992ms} (the present author also shared this belief in the past~\cite{Pedro:2016rih}[page 99]). Following this belief, the special role of the little group of rotations and the time evolution in our definition of a (special relativistic) Quantum Field Theory seems to be a step back in the road towards a general relativistic quantum theory. Thus, our definition would not be of much value since a rigorous definition of a Quantum Yang-Mills theory should mark \emph{``a turning point in the mathematical understanding of quantum field theory, with a chance of opening new horizons for its applications''}, as stated in reference~\cite{prize}.

But in the Hamiltonian formalism (and so in any quantum theory), the diffeomorphisms are generated by constraints~\cite{Mukhanov:1994zn} while the Poincare transformations act in a non-trivial way in the algebra of operators. Hence, the diffeomorphisms cannot be a generalization of a non-trivial Poincare symmetry and so, \emph{a priori} diffeomorphisms are not incompatible with the special role of the little group of rotations and the time evolution (see reference~\cite{Mukhanov:1994zn} and Section~\ref{sec:gravity}). \emph{A priori}, the formalism of gauge-invariant Quantum Field Theory may be enough to define quantum gravity.




The crucial fact that allows us to distinguish the Poincare symmetry from diffeomorphisms and gauge symmetry is the fact that the time-evolution generated by the Hamiltonian is \emph{not} a diffeomorphism or a gauge symmetry. There are indeed diffeomorphisms which advance the time coordinate of the phase-space (as we would expect from the time-evolution), however they differ from the time-evolution in how they affect the wave-function as a whole.
Thus, the Poincare symmetry is not a diffeomorphism or a gauge symmetry and it is not generated by constraints. The algebra of operators is a non-trivial representation of the Poincare group, while it is diffeomorphism invariant and gauge invariant. The generators of the (global) Poincare group are also diffeomorphism invariant and gauge invariant.

We define mathematically a general relativistic Quantum Field Theory, verifying all the properties we expect from Quantum Gravity, without adding complexity with respect to the mathematical definition of gauge-invariant Quantum Field Theory presented in the previous sections.

Unlike in Yang-Mills theory, the Hamiltonian of the Einstein-Cartan theory involves canonical commutation relations of analytic functions of operators, which is no challenge by itself when there is a non-polynomial function of only one of the operators involved~\cite{ccrfunctions}. With the Weyl ordering of products of operators and the measure defined in the previous section, there is no challenge or ambiguity if the kinetic part of the Hamiltonian is polynomial both in the fields and in its canonical conjugates while the potential is rational in the fields only (with no dependence on the canonical conjugates), just like in the Hydrogen atom in non-relativistic Quantum Mechanics. Note that since the Hamiltonian is defined in a 4-dimensional space-time, the diffeomorphisms (and all other constraints) are polynomial in the fields and its canonical conjugates, unlike in Loop Quantum Gravity~\cite{Thiemann:2007zz,Thiemann:2020cuq}.



Moreover, no vaccuum state needs to be defined (the Hamiltonian needs not to be bounded from below), thus there is no need for reguralization and so renormalization plays no essential role (at the non-perturbative level).

Moreover, the operator ordering ambiguity in Loop Quantum Gravity may limit it to the same peculiar notion of ``continuum'' through renormalization as in (Wilson's) Quantum Field Theory, as suggested in reference~\cite{Thiemann:2020cuq}:
``[...] the ultimate goal is to use Hamiltonian renormalisation to find a continuum theory for canonical Quantum Gravity. Here we can use the LQG [Loop Quantum Gravity] candidate as a starting point because it is rather far developed, but of course the flow scheme developed can be applied to any other canonical programme''. We propose instead that Quantum Gravity at all stages of the quantization programme, should be seen as a particular case of a (special relativistic) Quantum Field Theory defined in the continuum, which is a major simplification with respect to Loop Quantum Gravity.

Therefore, in the quantization due to time-evolution, there is no obvious advantage of trying to rewrite the Hamiltonian as a polynomial in the canonical operators, in contrast with Loop Quantum Gravity. We can still do it after the definition of the quantum theory~\cite{Katanaev:2005xd}, for the purpose of numerical computations in perturbation theory for instance. 

We will start by defining the Hamiltonian for Classical General Relativity within the formalism of Quantization due to the time-evolution. Just for the purpose of solving initial-value problems within the classical theory, it is already advantageous since it allows us to define an Hamiltonian which is diffeomorphism invariant (unlike in the ADM formalism~\cite{constraintalgebra,Deser:1976ay}), avoiding the technical complexity of the definition of chains of constraints~\cite{Kiriushcheva:2008sf}. Moreover, the low-energy limit of Quantum Gravity must anyway be studied and it should match the classical theory.

The special role of the little group of rotations and the time evolution in our definition of a (special relativistic) Quantum Field Theory seems at first sight to be a step back in the road towards a general relativistic quantum theory~\cite{Isham:1992ms}. the tetrad field (which is a kind of square root of the metric)~\cite{constraintalgebra}. Fortunately, the results of reference~\cite{constraintalgebra}

The classical Action for the Einstein–Cartan theory of gravitation is:
\begin{align}
S=&\int d^4 x \mathcal{L}\\
\mathcal{L}=&e R\\
e=&\det{e_{\mu}^{a}}\\
R=&g^{\mu \nu} R_{\mu \nu}\\
R_{\mu \nu}=&R^{\alpha}_{\mu \alpha \nu}\\
R^{\alpha}_{\beta \mu \nu}=&e^{\alpha}_{a} e_{\beta}^{b}\partial_{\mu}{\omega^{a}_{\nu b}}
-e^{\alpha}_{a} e_{\beta}^{b}\partial_{\nu}{\omega^{a}_{\mu b}}
+e^{\alpha}_{a} e_{\beta}^{b}\omega^{a}_{\mu c}\omega^{c}_{\nu b}
-e^{\alpha}_{a} e_{\beta}^{b}\omega^{a}_{\nu c}\omega^{c}_{\mu b}\\
\omega^{a}_{\mu b}=&1/2 (\eta^{a c} e^{\nu}_{c} \eta_{b k}\delta_{\mu}^{\alpha}\delta_{\nu}^{\beta}-e^{\nu}_{b}\delta_{\mu}^{\alpha}\delta_{\nu}^{\beta}\delta^{a}_{k}-\eta^{a c} e^{\alpha}_{c}e_{\mu}^{j}\eta_{j k}
e^{\beta}_{b})(\partial_{\alpha}{e_{\beta}^{k}}-\partial_{\beta}{e_{\alpha}^{k}})\\
g^{\mu \nu}=&e^{\mu}_{a} e^{\nu}_{b} \eta^{a b}
\end{align}
Where $a, b, c, j, k\in\{0,1,2,3\}$.

Then we integrate by parts in the variable $\partial_{\nu}{e_{\alpha}^{a}}$ and replace: 
\begin{align}
\partial_{\alpha}{e}=&-e e_{\mu}^{a}\partial_{\alpha}{e^{\mu}_{a}}\\
\partial_{\alpha}{e_{\mu}^{a}}=&-e_{\nu}^{a}\partial_{\alpha}{e^{\nu}_{b}} e_{\mu}^{b}
\end{align}

We also define:
\begin{align}
  \chi_{a}^{\ b}=&\delta_{a}^{\ b}+v_{a} v^{b}\\
  e_{a b c}=&\partial_{\alpha}{e^{\beta}\,_{b}} e^{\alpha}\,_{a} e_{\beta}\,^{k}\eta_{k c}\\
  E_{a b}=& \chi_{a}^{\ a_1} v^{c} \chi_{b}^{\ a_2} e_{a_1 c b}\\
  E_{a}=& \chi_{a}^{\ a_1} v^{c} v^{b} e_{a_1 c b}\\
  T_{a b c}=&e_{a b c}-e_{b a c}\\
  T_{a b}=& v^{c} \chi_{a}^{\ a_1} T_{c a_1 b}\\
  A_{a b}=& \chi_{a}^{\ a_1} \chi_{b}^{\ a_2} (T_{a_1 a_2}-T_{a_2 a_1})\\
  T=&\chi_{a}^{\ a_1} \chi_{b}^{\ a_2} T_{a_1 a_2} \eta^{a b}\\
  S_{a b}=&\chi_{a}^{\ a_1}\chi_{b}^{\ a_2}(T_{a_1 a_2}+T_{a_2 a_1}-\frac{2}{3}\eta_{a_1 a_2} T)\\
  T_{a}=&v^c \chi_{a}^{\ a_2} v^{b} T_{c a_2 b}\\
  \mathcal{T}_{a b c}=& \chi_{a}^{\ a_1} \chi_{b}^{\ a_2} \chi_{c}^{\ a_3} T_{a_1 a_2 a_3}\\
  \mathcal{T}_{a b}=& \chi_{a}^{\ a_1} \chi_{b}^{\ a_2} v^{c} T_{a_1 a_2 c}
\end{align}
Where $v^a=v^\mu e_\mu^a$ are the tetrad components of the globally defined time-like vector verifying $v^\mu v_\mu=-1$,
$a,b,c,a_1,a_2,a_3\in\{0,1,2,3\}$ and $T_{c b a}=-T_{c a b}$.

A divergence term in the Lagrangian density does not contribute to the Action, since the fields have a constant value at infinity (see Section~\ref{sec:measure}). Up to a divergence term, the Lagrangian density is given by:
\begin{align}
  &\mathcal{L}\approx e\left(T_{a b}^{\ \ b} T^{a c}_{\ \ c} - \frac{1}{2} T_{a b c} T^{a c b} - \frac{1}{4}T_{a b c} T^{a b c}\right)=\\
  &=\frac{1}{4}S_{a b} S^{a b} e - \frac{2}{3}(T)^2 e - \frac{1}{2}A_{a b} \mathcal{T}^{a b} e 
  -2 \mathcal{T}^{a b}_{\ \ a} T_{b} e+\\ 
  &- \frac{1}{2}\mathcal{T}_{a b c} \mathcal{T}^{a c b} e - \frac{1}{4} \mathcal{T}_{a b c} T^{a b c}e+\mathcal{T}_{b a}^{\ \ a} \mathcal{T}^{b c}_{\ \ c} e+\frac{1}{4}\mathcal{T}_{a b} \mathcal{T}^{a b} e \nonumber
\end{align}

This is in agreement with reference~\cite[See eq. (9.17), with $T_{a b c}$ replaced by 
$T_{c b a}$ due to a different notation]{aldrovandi2012teleparallel}, up to a global sign of the Lagrangian density (which does not affect the classical equations of motion).
Then we vary the resulting Lagrangian density in the variables
$\partial_{\alpha}{e^{\beta}_{a}}$, to obtain the polymomentum~\cite{polymomentum}:
\begin{align}
&p^{a b}=\frac{\delta\mathcal{L}}{\delta \partial_{\alpha}{e^{\beta}_{a}}}v_{\alpha} e^{\beta}_{c}\eta^{c b}=\\
&=\bigl(\frac{\delta\mathcal{L}}{\delta S_{a_3 a_4}}\frac{\delta S_{a_3 a_4}}{\delta T_{a_1 a_2}}
+\frac{\delta\mathcal{L}}{\delta T}\frac{\delta T}{\delta T_{a_1 a_2}}
+\frac{\delta\mathcal{L}}{\delta A_{a_3 a_4}}\frac{\delta A_{a_3 a_4}}{\delta T_{a_1 a_2}}
+\frac{\delta\mathcal{L}}{\delta T_{a_3}}\frac{\delta T_{a_3}}{\delta T_{a_1 a_2}}\bigr)
\frac{\delta T_{a_1 a_2}}{\delta \partial_{\alpha}{e^{\beta}_{a}}} v_{\alpha}e^{\beta}_{c}\eta^{c b}\nonumber\\
&=e\biggl(\frac{1}{2}S_{a_3 a_4}(\eta^{a_3 a_1}\eta^{a_4 a_2}+\eta^{a_4 a_1}\eta^{a_3 a_2}-\frac{2}{3}\eta^{a_3 a_4}\eta^{a_1 a_2})
-\frac{4}{3}T \eta^{a_1 a_2}+\nonumber\\
&-\frac{1}{2} \mathcal{T}_{a_3 a_4} (\eta^{a_3 a_1}\eta^{a_4 a_2}
-\eta^{a_4 a_1}\eta^{a_3 a_2})+2 \mathcal{T}_{a_3 a_4}\,^{a_4} \eta^{a_3 a_1} v^{a_2}\biggr)
\delta_{a_1}^{\ a}\delta_{a_2}^{\ b}\\
&=e\biggl(S^{a b}
-\frac{4}{3}T \eta^{a b}-\mathcal{T}^{a b}+2 \mathcal{T}^{a c}_{\ \ c} v^{b}\biggr)\\
&\mathcal{A}^{a b}=\chi^{a}_{\ a_1} \chi^{b}_{\ a_2}(p^{a_1 a_2}-p^{a_2 a_1})=
-2 e \mathcal{T}^{a b}\\
&\mathcal{P}=\eta_{a b}\chi^{a}_{\ a_1} \chi^{b}_{\ a_2} p^{a_1 a_2}=-4 e T\\
&\mathcal{S}^{a b}=\chi^{a}_{\ a_1} \chi^{b}_{\ a_2}
(p^{a_1 a_2}+p^{a_2 a_1}-\frac{2}{3}\eta^{a_1 a_2} \mathcal{P})=2 e S^{a b}\\
&p^{a}=v_b p^{a b}=2 e \mathcal{T}^{a c}_{\ \ c}
\end{align}

The Hamiltonian density in 3-dimensional space is: 

\begin{align}
&\mathcal{H}=v_{\alpha}\frac{\delta\mathcal{L}}{\delta \partial_{\alpha}{e^{\beta}_{a}}}v^{\mu}\partial_{\mu} e^{\beta}_{a}-\mathcal{L}=
p^{a b}T_{a b}+p^{a b}E_{a b}-\mathcal{L}=-\mathcal{L}+\frac{1}{4}\mathcal{A}^{a b} A_{a b}+\frac{1}{4}\mathcal{S}^{a b}S_{a b}+\nonumber\\
&+\frac{1}{3}\mathcal{P} T+p^{a} T_{a}+\frac{1}{2}\mathcal{A}^{a b} E_{a b}+\frac{1}{2}\mathcal{S}^{a b}E_{a b}
+\frac{1}{3}\mathcal{P} E_{a}\,^{a}+p^{a} E_{a}\\
&\approx e\biggl(\frac{1}{4}S^{a b} S_{a b} - \frac{2}{3}(T)^2
+S^{a b}E_{a b}-\frac{4}{3}T E_{a}\,^{a}-\mathcal{T}^{a b} E_{a b} +2 \mathcal{T}^{a b}\,_b E_{a}+\nonumber\\ 
&+ \frac{1}{2}\mathcal{T}_{a b c} \mathcal{T}^{a c b} + \frac{1}{4} \mathcal{T}_{a b c} T^{a b c}-\mathcal{T}_{b a}^{\ \ a} \mathcal{T}^{b c}_{\ \ c}-\frac{1}{4}\mathcal{T}_{a b} \mathcal{T}^{a b}\biggr)
\end{align}

As expected, the Hamiltonian density in 3-dimensional space only depends on the momenta $\mathcal{S}^{m n}$ and $\mathcal{P}$ through $S^{m n}$ and $T_{0}$:

\begin{align}
  &\mathcal{H}=\frac{1}{16 e}\mathcal{S}^{a b} \mathcal{S}_{a b} - \frac{1}{24 e} (\mathcal{P})^2
  +\frac{1}{2}\mathcal{S}^{a b} E_{a b}
  +\frac{1}{3}\mathcal{P} E_{a}\,^{a}-e\biggl(\mathcal{T}^{a b} E_{a b}+2 \mathcal{T}^{a b}\,_{b} E_{a}+\nonumber\\ 
  &+\frac{1}{2}\mathcal{T}_{a b c} \mathcal{T}^{a c b} + \frac{1}{4} \mathcal{T}_{a b c} T^{a b c}-\mathcal{T}_{b a}^{\ \ a} \mathcal{T}^{b c}_{\ \ c}-\frac{1}{4}\mathcal{T}_{a b} \mathcal{T}^{a b}\biggr)
\end{align}

The Hamiltonian density in 4-dimensional space is given by:
\begin{align}
&\mathcal{H_4}=-p^{a b}T_{a b}-p^{a b}E_{a b}+\mathcal{H}=\\
&=\frac{1}{4}Y_{a b} \eta^{c a} \eta^{j b} \theta_{c j}+\frac{1}{4}Z_{a b} \eta^{c a} \eta^{j b} \kappa_{c j}+\frac{1}{3}W X+W_{a} X_{b} \eta^{b a}+\\
&- \frac{1}{16}Z_{a b} Z_{c j} \eta^{a c} \eta^{b j} {e}^{-1}+\frac{1}{24}X X {e}^{-1} - \frac{1}{2}\eta^{a b} \eta^{c j} \eta^{k l} \varepsilon_{a c k} \varepsilon_{b l j} e
-\eta^{a b} \eta^{c j} \eta^{k l} \varepsilon_{a c b} \varepsilon_{j k l} e+\\
&+\frac{1}{4}\eta^{a b} \eta^{c j} \varepsilon_{a c} \varepsilon_{b j} e - \frac{1}{4}\eta^{a b} \eta^{c j} \eta^{k l} \varepsilon_{a c k} \varepsilon_{b j l} e
\end{align}
  
The constraints are the diffeomorphisms, local Lorentz transformations and global translations:
  \begin{align}
   G_\mu=&p^{\nu}_{a} \partial_{\mu} e_{\nu}^{a}-\partial_{\nu}(p^{\nu}_{a} e_{\mu}^{a})+p^{\nu}_{v} \partial_{\mu} v_{\nu}-\partial_{\nu}(p^{\nu}_{v} v_{\mu})\\
   J_{a b}=&p^{\mu}_{a} e_{\mu}^{c}\eta_{c b}-p^{\mu}_{b} e_{\mu}^{c}\eta_{c a}\\
   P_\mu=&\int d^4x \biggl(p^{\nu}_{a} \partial_{\mu} e_{\nu}^{a}+p^{\nu}_{v} \partial_{\mu} v_{\nu}\biggr)
  \end{align}

Note that the diffeomorphisms and global translations are consistent with each other,
but they are two independent constraints. The BRST charge is:
\begin{align}
 G=& p_{\nu}^{a}c^{\mu}\partial_{\mu}e^{\nu}_{a}-p_{\nu}^{a}e^{\mu}_{a}\partial_{\mu}c^{\nu}
 +\pi_{\nu}c^{\mu}\partial_{\mu}v^{\nu}-\pi_{\nu}v^{\mu}\partial_{\mu}c^{\nu}
 +i \partial_{\beta}c^{\alpha} c^{\beta}b_{\alpha}
\end{align}

\section{Conclusions}
\label{sec:conclusion}

Using the fact that there is a wave-function associated to any
probability distribution, we study a class of statistical field
theories in four-dimensional space-time where the (classical)
canonical coordinates when modified by the unitary time evolution (of the type of non-relativistic Quantum Mechanics), verify the canonical commutation relations. We then extend these statistical field theories to include non-trivial gauge symmetries and show that these theories have all the features of a gauge-invariant relativistic quantum field theory in four-dimensional space-time.


\addcontentsline{toc}{section}{References}
\footnotesize
\singlespacing 
\bibliography{Poincare}{}
\bibliographystyle{utphysMM}
\begin{acronym}[nuMSM]
\acro{2HDM}{two-Higgs-doublet model}
\acro{ATLAS}{A Toroidal LHC ApparatuS}
\acro{BR}{Branching Ratio}
\acro{BGL}{Branco\textendash{}Grimus\textendash{}Lavoura}
\acro{BSM}{Beyond the Standard Model}
\acro{CL}{Confidence Level}
\acro{cLFV}{charged Lepton Flavor Violation}
\acro{CLIC}{Compact Linear Collider}
\acro{CMS}{Compact Muon Solenoid}
\acro{CP}{Charge-Parity}
\acro{CPT}{Charge-Parity-Time reversal}
\acro{DM}{Darkmatter}
\acro{EDM}{Electric Dipole Moment}
\acro{EFT}{Effective Field Theory}
\acro{EW}{Electroweak}
\acro{EWSB}{Electroweak symmetry breaking}
\acro{FCNC}{Flavour Changing Neutral Current}
\acro{MET}{Missing Transverse Energy}
\acro{MFV2}{Minimal Flavor Violation with two spurions}
\acro{MFV6}{Minimal Flavor Violation with six spurions}
\acro{GIM}{Glashow\textendash{}Iliopoulos\textendash{}Maiani}
\acro{GNS}{Gelfand-Naimark-Segal}
\acro{GUT}{Grand unified theory}
\acro{ILC}{International linear collider}
\acro{LEP}{Large electron\textendash{}positron collider}
\acro{LFC}{Lepton flavor conservation}
\acro{LFV}{Lepton Flavor Violation}
\acro{LHC}{Large Hadron Collider}
\acro{MFV}{Minimal flavour violation}
\acro{MIA}{Mass insertion approximation}
\acro{MSSM}{Minimal Supersymmetry Standard Model}
\acro{nuMSM}[$\nu$MSM]{minimal extension of the Standard Model by three right-handed neutrinos}
\acro{PS}{Pati-Salam}
\acro{PT}[$\mathrm{p_T}$]{transverse momentum}
\acro{QCD}{Quantum chromodynamics}
\acro{RG}{Renormalization group}
\acro{RGE}{Renormalization group equation}
\acro{SM}{Standard Model}
\acro{SUSY}{Supersymmetry, Supersymmetric}
\acro{VEV}{Vacuum expectation value}
\acro{MEG}{Muon to electron and gamma}
\acro{NP}{New Physics}
\acro{NH}{Normal hierarchy}
\acro{IH}{Inverted hierarchy}
\acro{CKM}{Cabibbo\textendash{}Kobayashi\textendash{}Maskawa}
\acro{PMNS}{Pontecorvo-Maki-Nakagawa-Sakata}

\end{acronym}
\normalsize
\onehalfspacing 
\end{document}